\documentclass{article}

\usepackage{arxiv}

\usepackage[utf8]{inputenc} 
\usepackage[T1]{fontenc}    
\usepackage[hidelinks]{hyperref}       
\usepackage{url}            
\usepackage{booktabs}       
\usepackage{amsfonts}       
\usepackage{nicefrac}       
\usepackage{microtype}      
\usepackage{cleveref}       
\usepackage{graphicx}
\usepackage[square, sort&compress, numbers]{natbib}
\usepackage{doi}

\title{Generating Uniform Magnetic Fields Using Pulsed Field Magnetisation}

\date{February 9, 2024}

\author{ \href{https://orcid.org/0000-0002-7399-6166}{\includegraphics[scale=0.06]{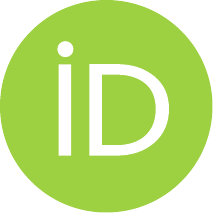}\hspace{1mm}Dian Weerakonda} \\
	Department of Engineering\\
	University of Cambridge\\
	Cambridge, CB2 1PZ \\
    U.K. \\
	\texttt{dw620@cam.ac.uk} \\
	\And
	\href{https://orcid.org/0000-0003-1574-9543}{\includegraphics[scale=0.06]{orcid.pdf}\hspace{1mm}Benjamin Bryant} \\
	Oxford Instruments NanoScience\\
	Abingdon\\
	Oxfordshire, OX13 5QX \\
    U.K. \\
	\texttt{ben.bryant@oxinst.com} \\
	\And
	\href{https://orcid.org/0000-0003-4962-7149}{\includegraphics[scale=0.06]{orcid.pdf}\hspace{1mm}Anthony Dennis} \\
	Department of Engineering\\
	University of Cambridge\\
	Cambridge, CB2 1PZ \\
    U.K. \\
	\texttt{ad466@cam.ac.uk} \\
    \And
	\href{https://orcid.org/0000-0003-0622-9298}{\includegraphics[scale=0.06]{orcid.pdf}\hspace{1mm}Tobia Nava} \\
	Department of Engineering\\
	University of Cambridge\\
	Cambridge, CB2 1PZ \\
    U.K. \\
	\texttt{tsn23@cam.ac.uk} \\
    \And
	\href{https://orcid.org/0000-0003-0712-3102}{\includegraphics[scale=0.06]{orcid.pdf}\hspace{1mm}John Durrell} \\
	Department of Engineering\\
	University of Cambridge\\
	Cambridge, CB2 1PZ \\
    U.K. \\
	\texttt{jhd25@cam.ac.uk} \\
}

\renewcommand{\shorttitle}{Uniform Fields with PFM}

\hypersetup{
pdftitle={Generating Uniform Magnetic Fields - Preprint},
pdfsubject={},
pdfauthor={Dian Weerakonda},
pdfkeywords={arxiv, PFM, preprint},
}

\begin{document}
\maketitle

\begin{abstract}
	Bulk high-temperature superconductors (HTS) are capable of generating very strong magnetic fields while maintaining a relatively compact form factor. Solenoids constructed using stacks of ring-shaped bulk HTS have been demonstrated to be capable of nuclear magnetic resonance (NMR) spectroscopy and magnetic resonance imaging (MRI). However, these stacks were magnetised via field cooling (FC), which typically requires a secondary superconducting charging magnet capable of sustaining a high magnetic field for a long period. A more economical alternative to FC is pulsed field magnetisation, which can be carried out with a magnet wound from a normal conductor, such as copper. In this work, we present a technique we have developed for iteratively homogenising the magnetic field within a stack of ring-shaped bulk HTS by manipulating the spatial profile of the applied pulsed field.
\end{abstract}


\section{Introduction}
Bulk high-temperature superconductors (HTS) promise to replace permanent magnets in multiple applications where performance is directly correlated with the strength of the magnetic field~\cite{roadmap2018, ainslie2014, nakagawa2012}. Among these is nuclear magnetic resonance (NMR) spectroscopy. While NMR spectrometers typically make use of magnets made from low-temperature superconductors (LTS) operating above 9~T, permanent magnet-based benchtop NMR spectrometers are not uncommon. They require far less lab space, and the purchase and maintenance costs are only a fraction of that of their superconducting counterparts~\cite{vanbeek2021}. However, the sensitivity and resolution of these systems are limited by their modest magnetic fields, where the current state-of-the-art is unable to reach fields above 3~T~\cite{qmagnetics}.

On the other hand, single-grain (RE)Ba$_2$Cu$_3$O$_{7-\delta}$ (REBCO), where RE is a rare-earth element, ring-shaped bulks have been used by Nakamura \textit{et al.} to construct a bulk HTS magnet for NMR spectroscopy that was able to operate at 4.7~T~\cite{nakamura2015}. While the homogeneity of this magnet was much lower than that which typical commercial systems achieve, it demonstrated that bulk HTS has the potential to replace permanent magnets in benchtop NMR systems. However, perhaps the biggest downside to their implementation, from a practical standpoint, was the method of magnetisation---field cooling (FC). FC requires a secondary charging magnet that can fully envelop the magnet being charged, which, in their case, was a large-bore superconducting magnet. Should such a magnetising technique be implemented in a commercial system, the large charging magnet will need to be shipped with the spectrometer magnet for charging onsite. Alternatively, the spectrometer magnet will need to be magnetised in the factory and somehow kept at its operating temperature (< 95~K) during shipping and installation. Both of these methods pose significant difficulties for implementation that are not present if a pulsed field is used for magnetisation.

Pulsed field magnetisation (PFM) of bulk HTS can be performed by discharging a bank of capacitors into a magnet wound from a normal conductor, such as copper, to generate a strong magnetic field over a small period of time (<~200~ms)~\cite{mizutani1998, tsui2022}. The use of copper significantly reduces the manufacturing and operating costs of the charging magnet. Furthermore, the small footprint of the charging magnet, resulting from the absence of stringent cooling demands from the copper, allows the charging magnet to be integrated into the main spectrometer magnet assembly. In this work, we demonstrate that the spatial profile of the pulsed field applied to a solenoid constructed out of a stack of ring-shaped EuBCO bulks can be varied to iteratively modify (and homogenise) the trapped field.

\section{Experimental Method}
\subsection{Field Cooling}
Four EuBCO ring-shaped bulks with a 60~mm outer diameter, 36~mm inner diameter, and 17~mm high were stacked together to form the bulk HTS solenoid. Each bulk was fully encapsulated in a ring of copper with a 5~mm wall thickness to minimise any thermal gradients between different sections of the solenoid. This stack was then field-cooled at 77~K with an applied field of 3.5~T. Upon the removal of the applied field, a Hall probe mounted to a linear stage was used to map the $z$-component (the direction along the stack) of the magnetic field along the central cylindrical axis of the solenoid in 1~mm steps. The magnet used for FC has a sufficiently large uniform region that we may assume each of the four bulks was fully magnetised following the procedure.

\subsection{Pulsed Field Magnetisation}
\begin{figure}[h]
    \centering
    \includegraphics[width=0.65\textwidth]{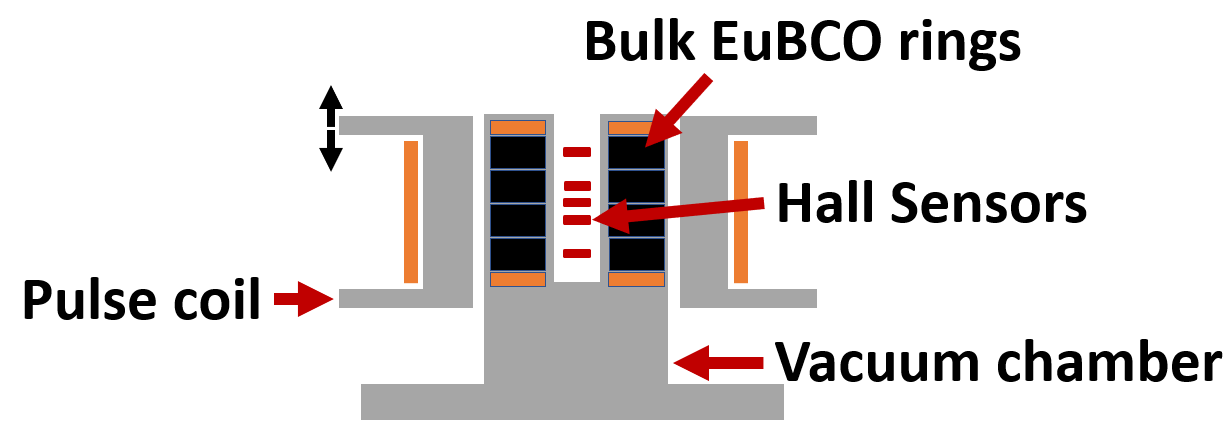}
    \caption{Pulsed field magnetisation setup. Four EuBCO rings are assembled inside a 316L stainless steel vacuum chamber with a room temperature bore, on the coldhead of a GM cryocooler. A liquid nitrogen-cooled solenoid, whose vertical position is adjustable, is placed around the vacuum chamber to generate the pulsed field.}
    \label{fig:setup}
\end{figure}

The setup used for PFM is shown in Fig.~\ref{fig:setup}. The bulks were assembled on the coldhead of a Gifford-McMahon cryocooler inside a 316L stainless steel vacuum chamber with a room temperature bore going through the centre of the stack. Heaters were used at the top and bottom of the stack to control the temperature. To generate the pulsed field, a copper solenoid connected to a bank of capacitors was used, with an IGBT acting as a switch that connects the capacitors to the solenoid. Details of this system can be found in~\cite{tsui2022w}. The solenoid was placed on a translating table such that its vertical position could be adjusted as required to modify the spatial profile of the pulse applied to the stack. During operation, the copper solenoid was cooled with liquid nitrogen to minimise its electrical resistance and hence maximise efficiency. Five Hall sensors were placed inside the room temperature bore to monitor the magnetic field during and after the pulse application.

Starting with an initial pulse with the centre of the coil aligned with bottom bulk, pulses with peak field values in the range of 1.3-1.7~T were applied with the pulse coil aligned at various positions of the stack to minimise the trapped magnetic field gradient along the $z$-axis of the stack. This allowed the trapped field within the stack to be iteratively homogenised, with only a small delay (5~minutes) between subsequent pulses to allow for thermal gradients induced by the preceding pulse to dissipate. Once the process of iterative homogenisation was completed, the Hall sensors inside the bore were removed and a single Hall sensor mounted to a linear stage was used to measure the $z$-component of the trapped magnetic field in the stack along the central cylindrical axis in 1~mm steps.

\section{Results \& Discussion}
\begin{figure}
    \centering
    \includegraphics[width=0.42\textwidth]{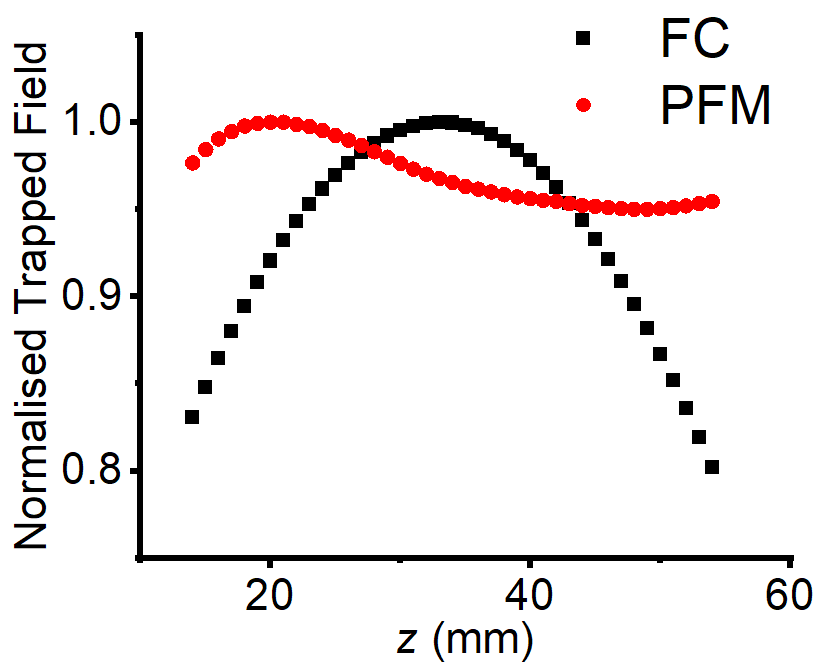}
    \caption{Normalised axial magnetic field profiles obtained with field cooling and PFM using a Hall sensor attached to a linear stage. The central field with FC was 2.0~T and the central field with PFM was 0.3~T. The field variation across the length of the stack is 4$\times$ lower in the iteratively homogenised PFM case.}
    \label{fig:results}
\end{figure}

The normalised axial profiles obtained with field cooling and PFM are shown in Fig.~\ref{fig:results}. The central field achieved with FC is 2.0~T, while with PFM, it is 0.30~T. However, the total variation in field across the length of the solenoid in the PFM case is 4$\times$ lower. This variant of PFM, where the spatial profile of the trapped field is modified iteratively with a very small delay between iterations, could be highly instrumental when magnetising bulk HTS solenoids for applications where the uniformity of the field is crucial. 

Further development of this technique should be towards developing an automated routine incorporating high-resolution magnetic field maps to determine the exact spatial profile of the pulse required to shim the trapped field. Currently, there is only one degree of freedom for the spatial profile of the pulse, but, for example, a segmented pulse coil where each segment can be activated independently will allow the span of the pulse to also be controlled, which should allow for more precise tuning of the trapped field.

\section{Conclusion}
The ability of REBCO bulks to trap strong magnetic fields at relatively high temperatures has been utilised in multiple applications. Our results demonstrate that the PFM technique has benefits beyond economy and footprint. PFM could be used for not just magnetisation, but also the shimming of bulk HTS magnets for NMR spectroscopy or MRI. Further development of the PFM technique to reach trapped fields comparable to that of FC could provide a compelling case for the adoption of PFM as the standard technique of magnetisation of bulk HTS for applications that require precise control of the generated field.

\section*{Acknowledgments}
This work was supported by funding from Oxford Instruments NanoScience and the Engineering and Physical Sciences Research Council. This work was also supported by the Engineering and Physical Sciences Research Council’s Cambridge Royce facilities grant EP/P024947/1 and Sir Henry Royce Institute~--~recurrent grant EP/R00661X/1.

\bibliographystyle{IEEEtran}

\end{document}